\documentclass[oneside,onecolumn,9pt]{article}

\usepackage{xcolor}
\usepackage{graphicx}
\usepackage{authblk}

\title{Impact of interface defects on the band alignment and performance of TiO$_2$/MAPI/Cu$_2$O perovskite solar cells}

\author[1,2,3]{Nicolae Filipoiu}
\author[2]{Marina Cuzminschi}
\author[1,2,3]{Mihaela Cosinschi}
\author[1,2]{Calin-Andrei Pantis-Simut}
\author[4]{Kristinn Torfason}
\author[4]{Rachel~Elizabeth~Brophy}
\author[4]{Andrei Manolescu}
\author[3]{Roxana E. Patru}
\author[3]{Cristina Besleaga}
\author[3]{George Stan}
\author[3]{Ioana Pintilie}
\author[1,2,5]{George~Alexandru~Nemnes}

\affil[1]{Faculty of Physics, University of Bucharest, 077125 Magurele-Ilfov, Romania}
\affil[2]{Horia Hulubei National Institute for Physics and Nuclear Engineering, 077126 Magurele-Ilfov, Romania}
\affil[3]{National Institute of Materials Physics, Magurele 077125, Ilfov, Romania}
\affil[4]{Department of Engineering, School of Technology, Reykjavik University, Menntavegur~1, IS-102 Reykjavik, Iceland}
\affil[5]{Research Institute of the University of Bucharest (ICUB), 90 Panduri Street, 050663 Bucharest, Romania}

\date{}

\begin{document}

\maketitle


\begin{abstract}
Optimizing the interfaces in perovskite solar cells (PSCs) is essential
for enhancing their performance, improving their stability, and making
them commercially viable for large-scale deployment in solar energy
harvesting applications. Point defects, like vacancies, have a dual role,
as they can inherently provide a proper doping, but they can also reduce
the collected current by trap-assisted recombination. Moreover, they
can play an active role in ion migration and degradation. Using {\it ab
initio} density functional theory (DFT) calculations we investigate the
changes in the band alignment induced by interfacial vacancy defects in
a TiO$_2$/MAPI/Cu$_2$O based PSC. Depending on the type of the vacancy
(Ti, Cu, O, Pb, I) in the oxide and perovskite materials, additional
doping is superimposed on the already existing background. Their effect
on the performance of the PSCs becomes visible, as shown by SCAPS
simulations. The most significant impact is observed for $p$ type doping
of TiO$_2$ and $n$ type doping of Cu$_2$O, while the effective doping of
the perovskite layer affects one of the two interfaces. We discuss these
results based on modifications of the band structure near the active
interfaces and provide further insights concerning the optimization of
electron and hole collection.
\end{abstract}

\section{Introduction}

Perovskite solar cells (PSCs) captured a tremendous amount of attention in the last few years due to the rapid advancement in power conversion efficiency (PCE), nowadays reaching 27.3\%, which is larger than in silicon single crystal devices \cite{nrel}. The low fabrication costs of PSCs, based on chemical routes, is a major advantage compared to standard silicon photovoltaic technology, but issues related to stability and the optimization of the multi-layered structures for large area scalability still remain to be solved \cite{KHATOON2023437,Tao2023}. This requires choosing materials for electron and hole transporter layers (ETL, HTL) as well as adjustments to the perovskite absorber in order to get a proper band alignment at the active interfaces, taking into account the constraints concerning the fabrication process, like solvents, deposition temperature etc \cite{Lee_2022,SIN2023112054,JIAO2023101158}.

The stability of the PSCs depends on the ability to mitigate both intrinsic and extrinsic degradation mechanisms \cite{en14175431}. Whereas by encapsulation one can reduce the impact of moisture and oxygen, the intrinsic mechanism mainly derive from the ion migration in the perovskite (PRV) layer during the operation of the solar cell \cite{doi:10.1021/acs.jpclett.6b02375}. To contain the ion migration the structure of the ETL and HTL layers is essential. Although originally, Spiro-OMeTAD was used as HTL in the vast majority of PSC devices, the high costs and ion permeability focused the attention towards inorganic crystalline compounds, like Cu$_2$O, NiO etc \cite{https://doi.org/10.1002/nano.202000238,SAJID2023101378}. Most of these oxide materials can be processed as thin films by low temperature solution-based techniques or, alternatively, inverted PSC structures can be considered \cite{https://doi.org/10.1002/nano.202000200}. Typically, the ETL is less prone to ion migration, as it is usually formed by crystalline materials (e.g. TiO$_2$, ZnO, SnO$_2$), however, it plays an equally important role in the overall performance of the PSC. Under operating conditions, the voltage drops and the temperature increase typically enhance the ion migration, which may cause modifications of the I-V characteristics \cite{electronics14224428}.

Co-doping and intentional defect engineering of the ETL/HTL and of the perovskite absorber is essential for high-PCE and stable PSCs \cite{CAO2022420}. Usually, depending also on the synthesis route, these thin films already have a significant amount of unintentional point defects which tune the $p$/$n$ type character \cite{C5CC05205J,Srivastava2023,Euvrard2021} and may further facilitate ion migration. This is why an optimal amount of point-defects can ensure the proper band offsets at the interfaces, while preserving the ion-blocker quality. A larger amount of defects is typically found in and around the interface regions \cite{CHOUHAN2018150}, as the lattice mismatch between the two materials triggers more defects.   

The electronic properties of the two active interfaces, ETL-PRV and PRV-HTL, as ideal systems, have been investigated by {\it ab initio} calculations using density functional theory (DFT). Initially, the studies were mostly focused on the TiO$_2$-MAPI (methylammonium lead iodide) interface, analyzing the band-alignment and the influence of chlorine-iodine proportion \cite{doi:10.1021/jz501127k,doi:10.1021/jz501869f,C5CP05466D}, taking into account different possible terminations of the two material slabs \cite{Geng2016}, stability and charge separation \cite{YANG2018394,doi:10.1021/acs.jpclett.3c03536}. Subsequently, searching for higher mobility ETLs, two other candidate materials were investigated, ZnO \cite{NICOLAEV2016202,Si2017} and SnO$_2$ \cite{Sultana2019}, while other configurations, based on nanostructured TiO2/BaTiO3 electron transport bilayers were proposed \cite{MOHAMMADI2025118454}. At the same time, crystalline inorganic HTLs like Cu$_2$O have gained importance and {\it ab initio} studies revealed the importance of vacancies to reduce the trap states at the interface \cite{doi:10.1021/acsami.0c11187}. An overview of {\it ab initio} studies on perovskite materials and interfaces is found in Ref.\ \cite{MITRAN2023153}. However, due to the computational complexity, mostly ideal interfaces were considered.

On a macroscopic level, SCAPS simulation software \cite{BURGELMAN2000527} was widely used to analyze the influence of material parameters on the performance of the PSCs. Different configurations of the layers and materials have been investigated so far, amongst which we can mention the impact of work function of back contact \cite{MINEMOTO20141428}, design of lead-free flexible PSC \cite{Goje_2023}, an extensive study of several ETLs (PCBM, TiO$_2$, ZnO, SnO$_2$, IGZO, WS$_2$) and HTMs (Cu$_2$O, CuSCN, CuSbS$_2$, NiO etc) using combined DFT and SCAPS \cite{D2RA06734J,doi:10.1021/acs.energyfuels.3c00035,Baro2023}, PSCs with nitrogen-doped titanium dioxide as an inorganic hole transport layer \cite{inorganics11010003}, investigation of defects in SnX$_3$-based PSCs \cite{nano11051218}. Further performance optimizations were achieved using genetic algorithms \cite{HUNDE2023105420}. The impact of defects is implemented by additional interface defect layers (IDLs), as employed in the study of germanium-based PSCs \cite{MENG2022112427,ZHANG2023128084}. A recent review summarizes the advances in lead-free perovskite solar cell design using SCAPS simulations \cite{D5SU00526D}. Here, we use SCAPS to correlate defect-induced band structure modifications with the solar cell performance parameters.

In this paper we perform extensive investigations on the vacancy defects in the ETL, HTL and perovskite layers using DFT calculations on thin film interface structures. The additional doping introduced by vacancies is evidenced, which translate into band alignment modifications with respect to the reference PSC. This is shown for several disorder realizations indicating consistent trends. Considering as input the defect character, we assess the performance of the PSCs using SCAPS by introducing vacancy-specific IDLs at both ETL-PRV and PRV-HTL interfaces. An overview on the evolution of the relevant parameters of the solar cell, like the open-circuit voltage ($V_{\rm oc}$), short-circuit voltage ($I_{\rm sc}$) and fill factor (FF) is also provided.

\begin{figure}[t]%
\begin{center}
  \includegraphics[width=0.8\linewidth]{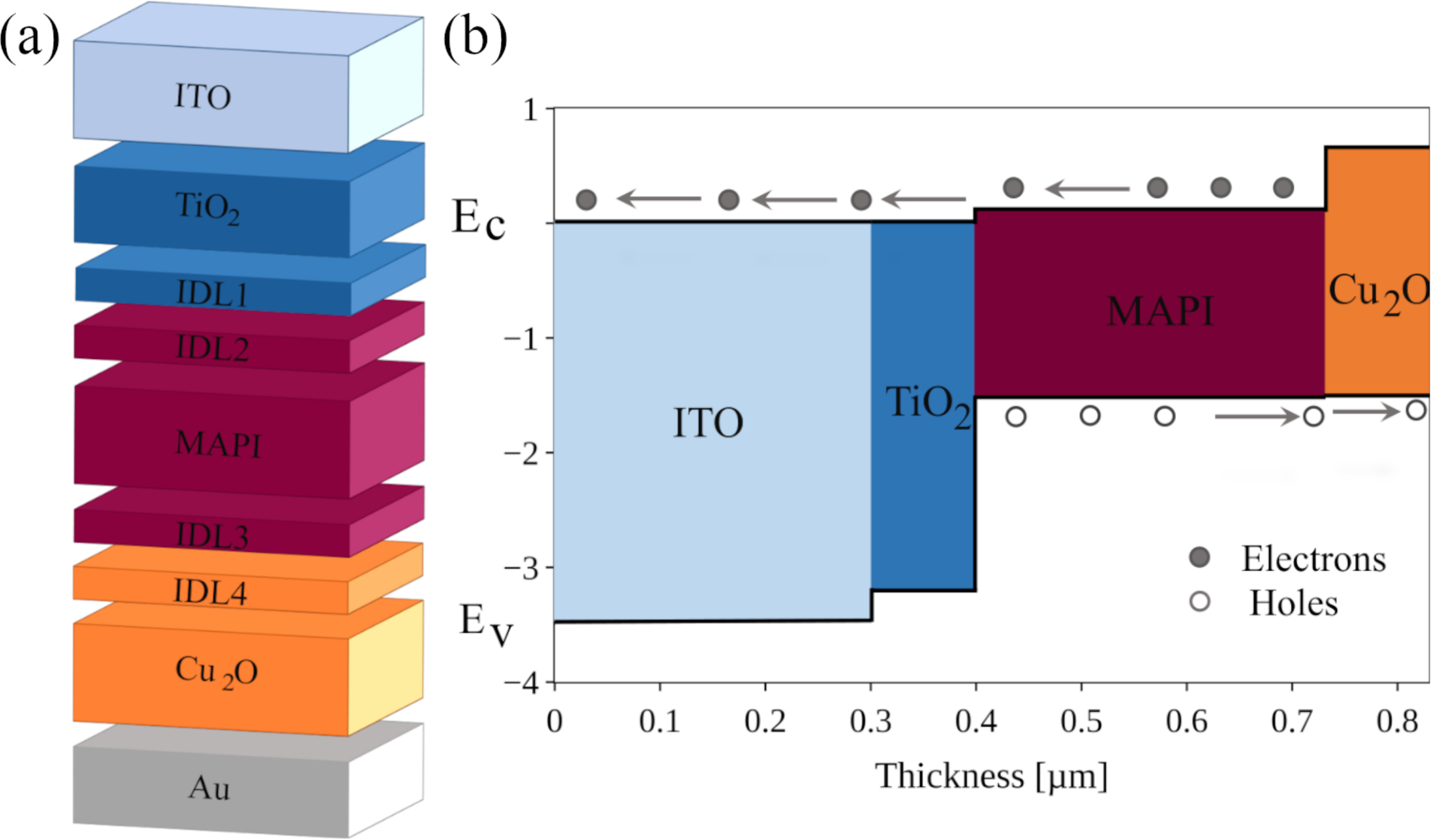}
\end{center}
	\caption{(a) The structure of the reference PSC: ITO as transparent conductive oxide, TiO$_2$ (ETL), MAPI perovskite absorber, Cu$_2$O (HTL), Au for the metallic back contact. The IDLs account for vacancies accumulations near the two interfaces. (b) A schematic band diagram for this structure evidences the small band offsets in the conduction band and valence band, for the ETL-PRV and PRV-HTL, respectively. }
    \label{refPSCstruct}
\end{figure}

\section{Computational Details}

\subsection{The reference PSC structure}

The reference PSC structure is depicted in Fig.\ \ref{refPSCstruct}, showing the five layers: ITO, TiO$_2$ (ETL), MAPI perovskite absorber, Cu$_2$O (HTL) and Au back contact. This is a rather typical configuration of a PSC with an inorganic HTL, which provides almost flat band condition with small band offsets at the ETL-PRV and PRV-HTL interfaces, for electrons and holes, respectively. It ensures the blocking of holes at the ETL and electrons at HTM, while the small energy differences in the conduction and valence bands, respectively, at the two interfaces enhance the collection of photogenerated carriers. However, small variations in the band structure arising at those interfaces, due to point defects (e.g. an excess of vacancies), can drastically alter the performance of the solar cell, as it will be shown in the following.

\subsection{{\it Ab initio} calculations for bulk materials and interface structures}

\begin{figure}[t]%
\begin{center}
 \includegraphics[width=0.80\linewidth]{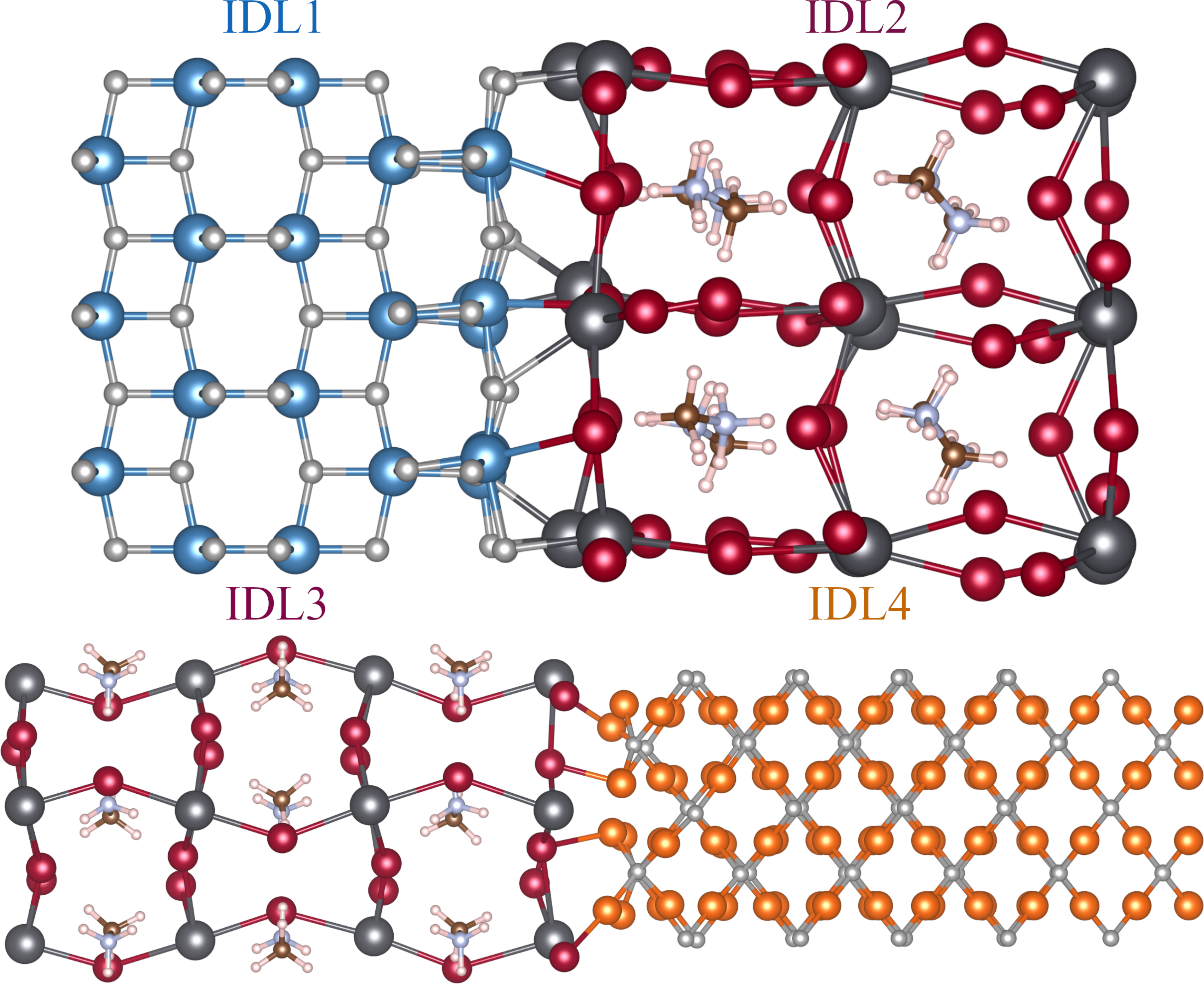}	
\end{center}
	\caption{Relaxed interface structures, MAPI@TiO$_2$, with Ti-O/Pb-I terminations and MAPI@Cu$_2$O, with Cu/Pb-I terminations. The supercells are periodic in the $x$-$y$ directions, while a 15 \AA\ vacuum was considered along the $z$-direction.}
    \label{MAPIaCu2O}
\end{figure}

The DFT calculations are implemented using SIESTA package \cite{0953-8984-14-11-302}, which employs a set of strictly localized numeric atomic orbitals that provides a linear scaling of the computational time with the system size. In the first step, the bulk materials are optimized using the local density approximation (LDA) in the parametrization of Ceperley and Alder \cite{PhysRevLett.45.566} and norm-conserving Troullier-Martins pseudopotentials \cite{PhysRevB.43.1993}. Throughout all calculations a double-$\zeta$ polarized (DZP) basis set and a meshcutoff of 300 Ry were used. For the sampling of the Brillouin zone, a Monkhorst-Pack grid of $5 \times 5 \times 5$ was employed. The relaxations were performed until the forces were less than 0.04 eV/\AA. In the case of Cu$_2$O we included LDA+U approximation for reproducing more accurately the experimental band gap, which is less demanding than using hybrid functionals \cite{YAN201434}, while for TiO$_2$ and MAPI it is reasonably reproduced without correction.

Using the optimized geometries for the unit cells, interface structures are assembled, as shown in Fig.\ \ref{MAPIaCu2O}. In order to reduce the mismatch strain as much as possible, the supercells are constructed as follows. For the interface $I_1$ (ETL-PRV) we considered a slab with 135 atoms for TiO$_2$ interfaced with 108 atoms for the tetragonal MAPI, ensuring a mismatch of less than 1.1\%. For the interface $I_2$ (PRV-HTL) we used  78 atoms for the orthorhombic MAPI phase on top of 140 atoms for the Cu$_2$O layer, with a mismatch smaller than 5\%. The difference between the bulk tetragonal and orthorhombic MAPI is minimal, and this choice reduces potential strain effects that can bring larger deviations in the band structure calculations. Moreover, MAPI often exhibits phase coexistence or dynamic transitions \cite{abia2022crystal,whitfield2016structures}. The completed structures have 243 atoms in the case of $I_1$ (MAPI@Cu$_2$O) and 218 atoms for $I_2$ (MAPI@Cu$_2$O). Given the large system sizes, a smaller meshcutoff of 150 Ry is enforced and, due to the lack of periodicity on the $z$-direction, a Monkhorst-Pack grid of $5 \times 5 \times 1$ was employed. 

Starting from the ideal interface structures, a number of 20 disorder realizations are constructed for each vacancy type (\textnumero Vacancies / \textnumero Atoms): (Interface $I_1$) $1/12=8.33\%$ $V_{\rm Pb}$, $2/32=6.25\%$ $V_{\rm I}$, $2/45= 4.44\%$ $V_{\rm Ti}$ and $4/90=4.44\%$ $V_{\rm O}$, (Interface $I_2$) $1/8=12.5\%$ $V_{\rm Pb}$, $1/22=4.55\%$ $V_{\rm I}$, $5/96=5.26\%$ $V_{\rm Cu}$ and $2/44=4.55\%$ $V_{\rm O}$. In spite of the fact that these are relatively high defect concentrations, inherently constrained by the limitations of the DFT calculations, the analysis provides important clues concerning the doping character and relative importance of the considered vacancy types. Moreover, one should note that a typical defect distribution would have a larger contribution near the interface compared to bulk of the crystalline structures, induced by the mismatch of the two materials.

For post-processing of the electronic structure calculations, including the projected density of states (PDOS), we used the SISL package \cite{zerothi_sisl}. The band alignment is determined from the PDOS of individual layers.

\begin{table*}[t]
\centering
\footnotesize
	\caption{SCAPS parameters used for the reference PSC. The IDLs are part of the corresponding oxide or MAPI layers. In IDLs, $(^*)$ marks the variation of the defect concentration between $10^{15}$ cm$^{-3}$ and $10^{18}$ cm$^{-3}$.}
\vspace*{0.5cm}
\hspace*{-3.5cm}
\begin{tabular} {l l l l l l l l l} 
\hline
Parameters             & ITO\cite{Baro2023} & TiO$_2$\cite{bransky1969hall,ait2022comparison,bansal2016evaluation,tan2016controllable} & IDL1 & IDL2 & MAPI\cite{Baro2023,abena2022numerical} & IDL3 & IDL4 & \hspace{-0.0cm}Cu$_2$O\cite{sajid2024boosting,sekkat2021open,pan2024high}\\  \hline     
Thickness \tiny{[nm]}         & 300 & 90 & 10 & 10 & 330 & 10 & 10 & 90\\
E$\rm{_g}$ \tiny{[eV]}          & 3.5 & 3.2 & 3.2 & 1.55 & 1.55& 1.55& 2.1 & 2.1\\
$\chi$ \tiny{[eV]}             & 4.0 & 4.0 & 4.0 & 3.9 & 3.9 & 3.9 & 3.2 & 3.2\\
$\varepsilon_r$          & 9 & 10 & 10 & 10 & 10 & 10 & 7.11 & 7.11\\
N$\rm{_c}$ \tiny{[cm$^{-3}$]}   & $2.2\times10^{18}$ &$2.0\times10^{18}$ & $2.0\times10^{18}$ & $2.75\times10^{18}$ & $2.75\times10^{18}$ & $2.75\times10^{18}$& $2.2\times10^{18}$& $2.2\times10^{18}$\\
N$\rm{_v}$ \tiny{[cm$^{-3}$]}      & $1.8\times10^{19}$ & $1.8\times10^{19}$ & $1.8\times10^{19}$ & $3.9\times10^{18}$ & $3.9\times10^{18}$ & $3.9\times10^{18}$ & $1.9\times10^{19}$ &$1.9\times10^{19}$\\
$\mu_n$ \tiny{[cm$^2$V$^{-1}$s$^{-1}$]}                   & 30 & 20 & 20 & 20 & 20 & 20 & 5 & 5\\
$\mu_p$ \tiny{[cm$^2$V$^{-1}$s$^{-1}$]}                   & 1 & 1 & 1 & 20 & 20 & 20 & 86 & 86\\
N$\rm{_D}$ \tiny{[cm$^{-3}$]}              & $1.0\times10^{19}$& $1.0\times10^{16}$& $1.0\times10^{16}$ & $1.0\times10^{16}$ & $1.0\times10^{10}$ & $1.0\times10^{10*}$ &$1.0\times10^{10*}$&$1.0\times10^{10}$\\
N$\rm{_A}$ \tiny{[cm$^{-3}$]}              &-&$1.0\times10^{10}$&$1.0\times10^{10*}$& $1.0\times10^{10*}$ & $1.0\times10^{10}$ & $1.0\times10^{16}$ & $1.0\times10^{17}$& $1.0\times10^{17}$\\
N$\rm{_t}$ \tiny{[cm$^{-3}$]}              &-&$1.0\times10^{13}$& $1.0\times10^{13}$ & $1.0\times10^{15}$ &$1.0\times10^{15}$ & $1.0\times10^{15}$ & $1.0\times10^{13}$ &$1.0\times10^{13}$\\
\hline
\end{tabular}
\label{SCAPS}
\end{table*}

\subsection{Performance analysis using SCAPS}

The impact of the different types of defects on the PSC performance is evaluated from the stationary J-V characteristics modelled by SCAPS. The program calculates the energy bands, carrier concentrations and currents for devices with up to 7 semiconductor layers, by solving the one-dimesional semiconductor equations together with the appropriate boundary conditions at the contacts \cite{BURGELMAN2000527}. Each layer is characterized by a series of parameters, like: band gap ($E_g$), affinity ($\chi$), relative dielectric permittivity ($\varepsilon_r$), conduction- and valence band effective density of states ($N_{\rm C}$ and $N_{\rm V}$), thermal velocities for electrons and holes ($v_{\rm th,n}$ and $v_{\rm th,p}$), electron and hole mobilities ($\mu_n$ and $\mu_p$), donor and acceptor concentrations ($N_{\rm D}$ and $N_{\rm A}$) and total concentration of defects $N_t$. A summary of parameters is shown in Table \ref{SCAPS}. In this configuration, the TiO$_2$ layer and the corresponding IDL1 are considered $n$-type, consistent with their role as the electron transport layer (ETL), where the electron mobility ($\mu_n$) and donor density ($N_{\rm D}$) dominate the charge transport properties and facilitate efficient electron extraction. Similarly, the Cu$_2$O layer and the corresponding IDL4 are considered $p$-type, reflecting their role as the hole transport layer (HTL), where the hole mobility ($\mu_p$) and acceptor density ($N_{\rm A}$) dominate the transport mechanism. Therefore, the defects play a dual role, of modifying band alignment via effective doping, while they act as recombination centers in the proximity of the interfaces.

Although the main MAPI absorber layer is considered intrinsic, the interface defect layers adjacent to the transport layers are designed to emulate interfacial band bending and additional recombination. Consequently, IDL2 exhibits ETL-like behavior through dominant donor-type characteristics, while IDL3 exhibits HTL-like behavior through dominant acceptor-type characteristics. Standard AM 1.5G illumination conditions were considered in all SCAPS simulations.

The defect concentrations of each IDL was varied independently between $10^{15}$ cm$^{-3}$ and $10^{18}$ cm$^{-3}$ in order to investigate the defect compensation effects for the base doping, which is donor-type for IDL1 and IDL2 and acceptor-type for IDL3 and IDL4. For the $I_1$ interface, IDL1 and IDL2 correspond to TiO$_2$ and MAPI, and they account for $V_{\rm Ti}$, $V_{\rm O}$ and $V_{\rm Pb}$, $V_{\rm I}$, respectively. Similarly, for the $I_2$ interface, we assign IDL3 and IDL4 to MAPI and Cu$_2$O defect layers, the latter modelling $V_{\rm Cu}$ and $V_{\rm O}$. The width of all IDLs is set to $l_d = 10$ nm. 
The performance of the PSCs is evaluated by analyzing the J-V characteristics, with a particular focus on key parameters such as the open-circuit voltage ($V_{\rm oc}$), short-circuit current density ($J_{\rm sc}$) and fill factor (FF). The variations in these parameters are correlated with the changes in the interfacial band alignment induced by the different types of vacancies, as revealed by the DFT calculations. This comprehensive analysis allows us to understand how specific defects at the interfaces can influence charge transport and recombination processes, ultimately affecting the overall efficiency of the solar cell.

\section{Results}
\label{ecmodels}

\subsection{Structural optimization and electronic structure of bulk TiO$_2$, Cu$_2$O and MAPI}

\begin{figure}%
\begin{center}
  \includegraphics[height=4.5cm]{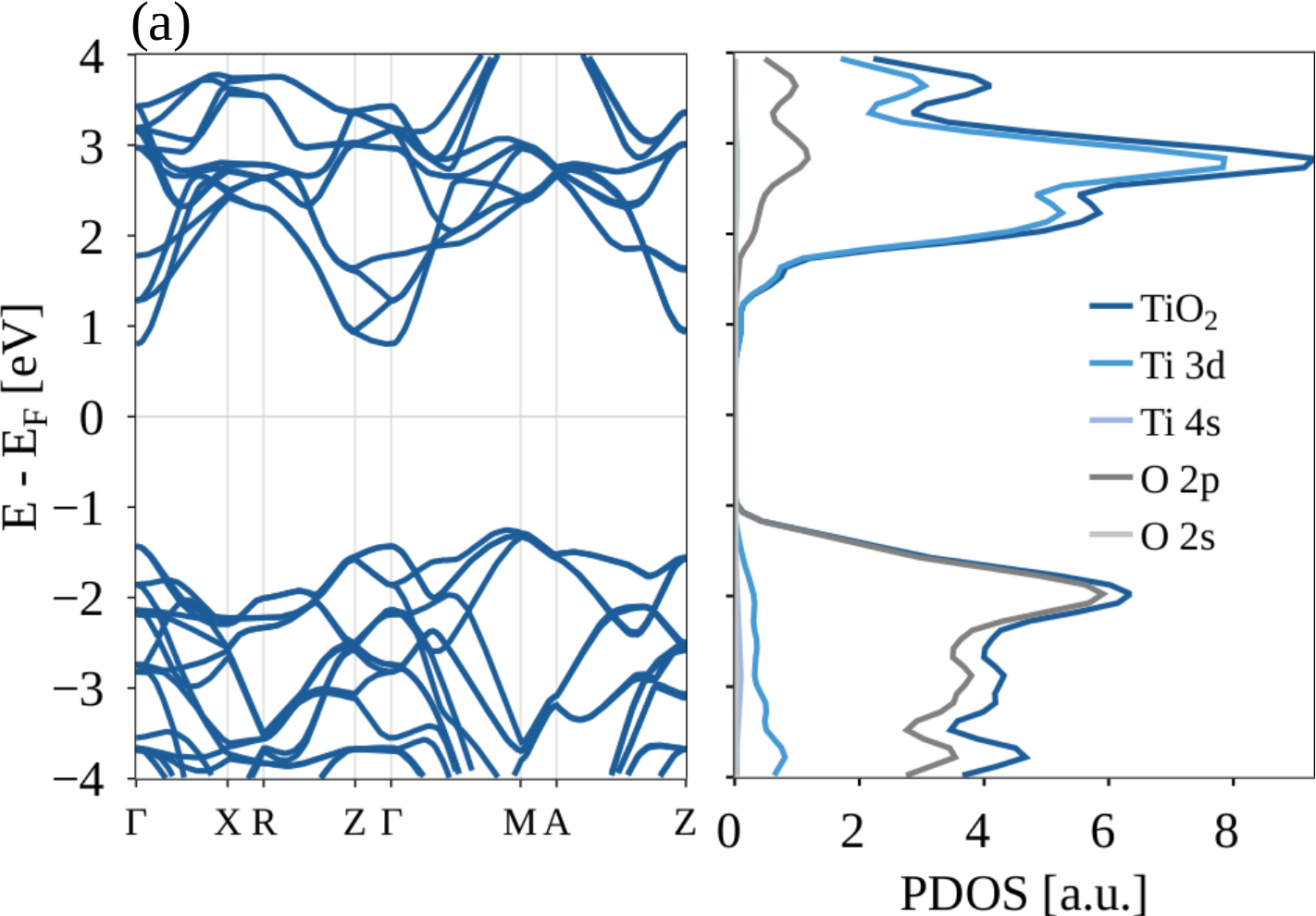}\hspace*{0.5cm}
  \includegraphics[height=4.5cm]{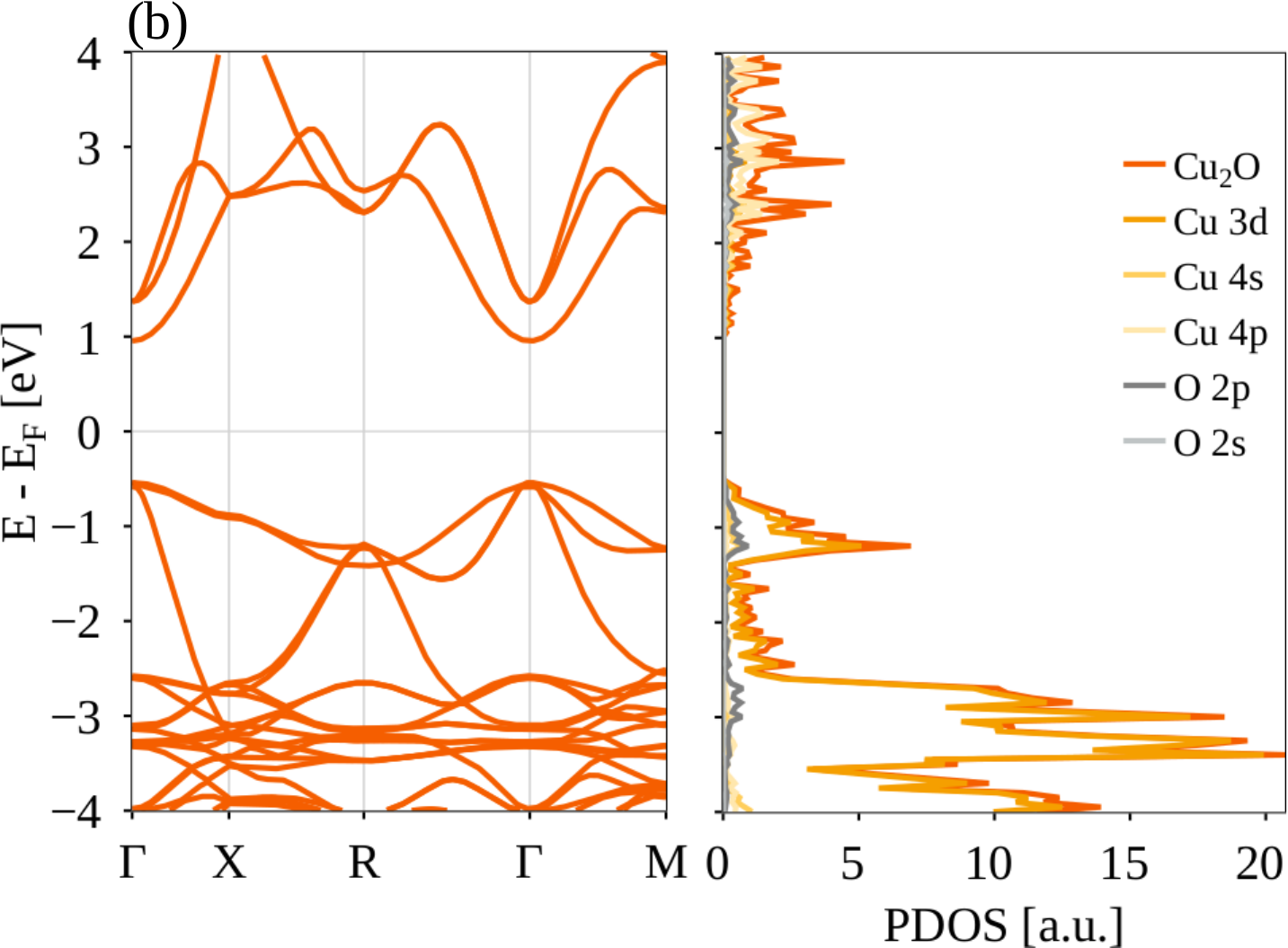}
\end{center}
	\caption{Band structure and PDOS for the ETL and HTM: (a) anatase-TiO$_2$, $E_g = 2.17$ eV; (b) cuprous oxide (Cu$_2$O), $E_g = 1.49$ eV, using a DFT+U correction with $U = 8$ eV and $J = 0.8$ eV on 3d Cu orbitals and $U = 12$ eV for 2p Oxygen orbitals. }
	\label{EH_bandstruct}
\end{figure}   

\begin{figure}%
    \begin{center}
      \includegraphics[height=4.5cm]{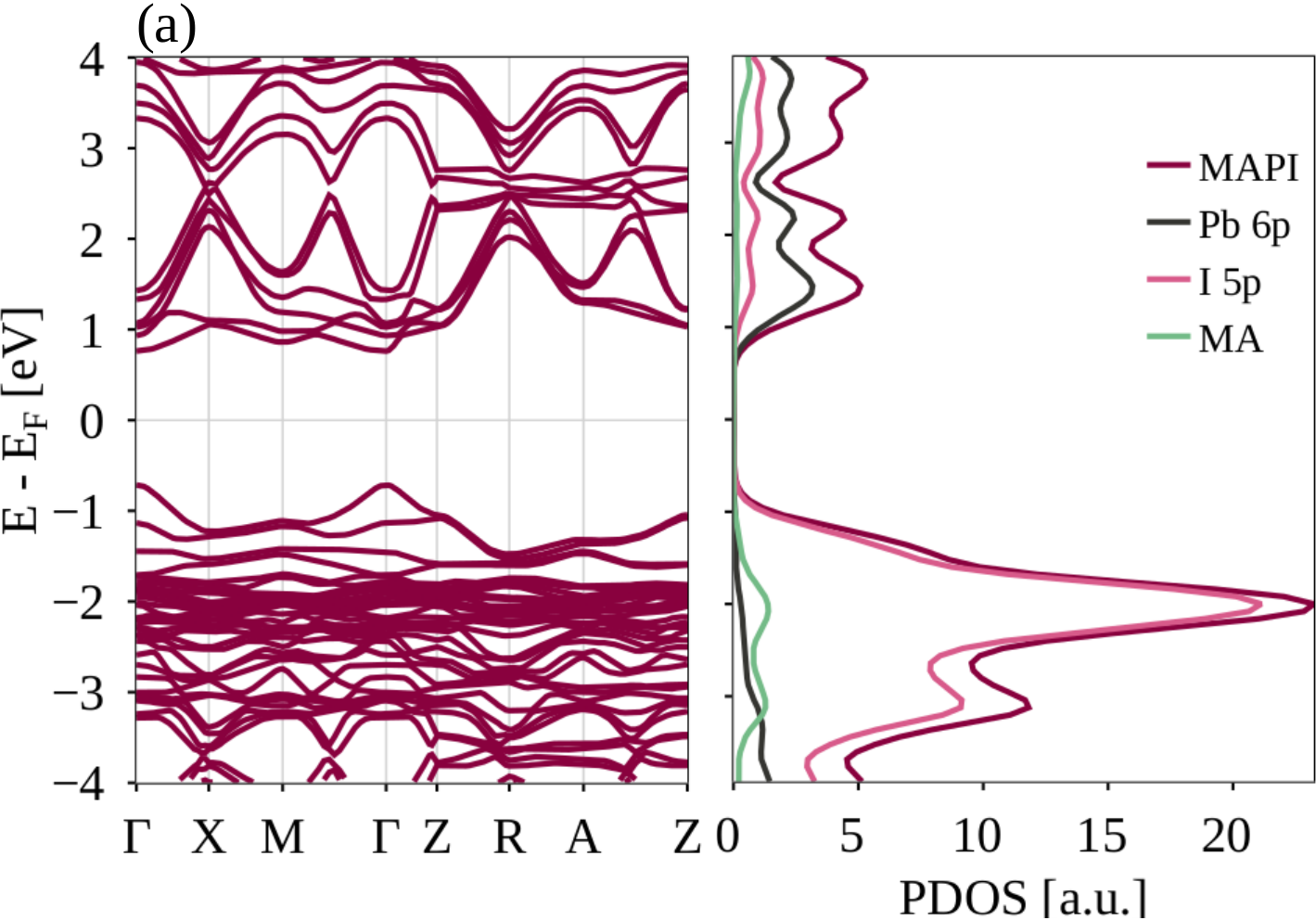}\hspace*{0.5cm}
      \includegraphics[height=4.5cm]{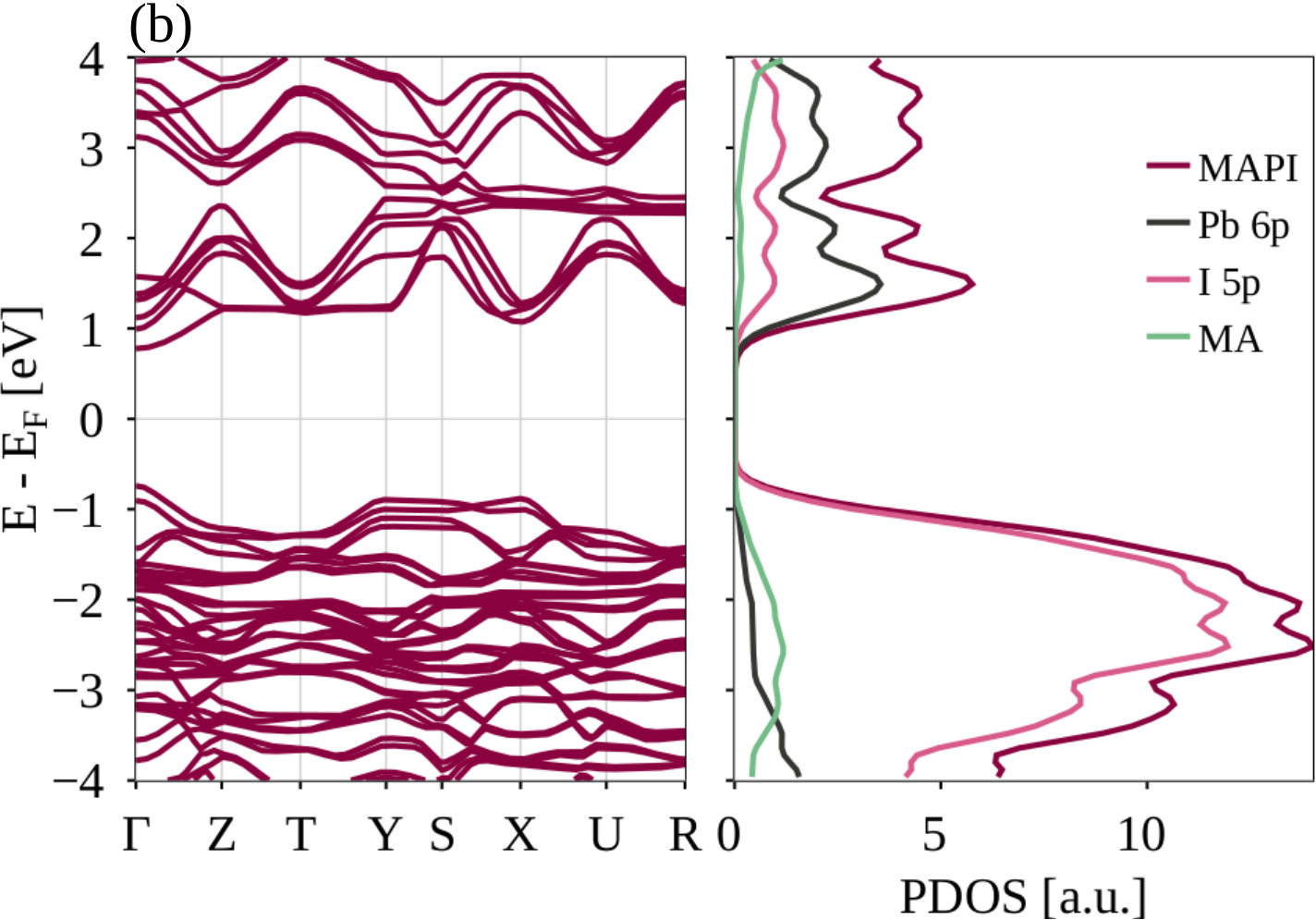}
    \end{center}
	\caption{Band structure and PDOS for MAPI: (a) tetragonal and (b) orthorhombic structures. The energy gap is similar in both instances, $E_g =1.48$ eV and $E_g =1.52$ eV, respectively.}
        \label{MAPI_tetr_bands}
\end{figure}

As part of the calibration process of the DFT calculations, we proceed with the structural optimization and electronic band structure calculations of bulk TiO$_2$, Cu$_2$O and MAPI materials. The atomic configurations are further used to optimize the interface structures.

Anatase-TiO$_2$ is commonly used as ETL in standard PSCs, in contrast to the high temperature phase, rutile-TiO$_2$. It is a relatively wide band gap semiconductor, with experimental $E_g = 3.22$ eV \cite{MA2021167107} and almost transparent for visible light. The unit cell contains 4 Ti atoms and 4 O atoms and the structure belongs to I4$_1$/amd space group. The band structure shown in Fig.\ \ref{EH_bandstruct}(a) indicates a direct band gap, with $E_g = 2.17$ eV. The edges of the conduction band (CB) and valence band (VB) mainly stem from the Ti-3d, Ti-4s, Ti-4p orbitals and O-2p orbitals, respectively.    

Cuprous oxide (Cu$_2$O) crystallizes in a cubic structure and belongs to the space group Pn$\bar{3}$m. This material has an experimental band gap value of approximately 2.1 eV \cite{PhysRevB.87.115111}, making it suitable for use as a semiconductor in various electronic devices and sensors \cite{10.1063/1.4919323}. Due to the limitations of DFT in accurately treating strongly correlated electron states, the band gap of Cu$_2$O is typically underestimated. To address this issue, the DFT+U correction was applied. As one can see from Fig.\ \ref{EH_bandstruct}(b), this correction increased the calculated band gap value to approximately 1.5 eV, bringing it closer to the experimental band gap. The PDOS analysis shows that the conduction and valence bands are mainly formed by Cu-4s and Cu-3d orbitals, respectively.

The MAPI perovskite is found in orthorhombic, tetragonal or cubic structures, with quite similar electronic properties. The ground state is orthorhombic, and it contains 48 atoms per unit cell. The experimental band gap of $\approx$ 1.6 eV is found with good accuracy from DFT calculation without DFT+U correction, as seen from Fig.\ \ref{MAPI_tetr_bands}, due to a fortouis cancellation between the underestimation provided by LDA and the overestimate caused by the lack of spin-orbit coupling (SOC) \cite{C4TA05033A}. The conduction band edge is composed of I-5p, while the top of the valence band is formed by Pb-6p orbitals. 

\subsection{Structural optimization and band alignment at ideal and defective TiO$_2$-MAPI and MAPI-Cu$_2$O interfaces}

    \begin{figure*}%
        \begin{center}
	  \hspace*{-4.7cm}(a) \hspace*{7.2cm}(b)\\
	  \includegraphics[width=0.47\linewidth]{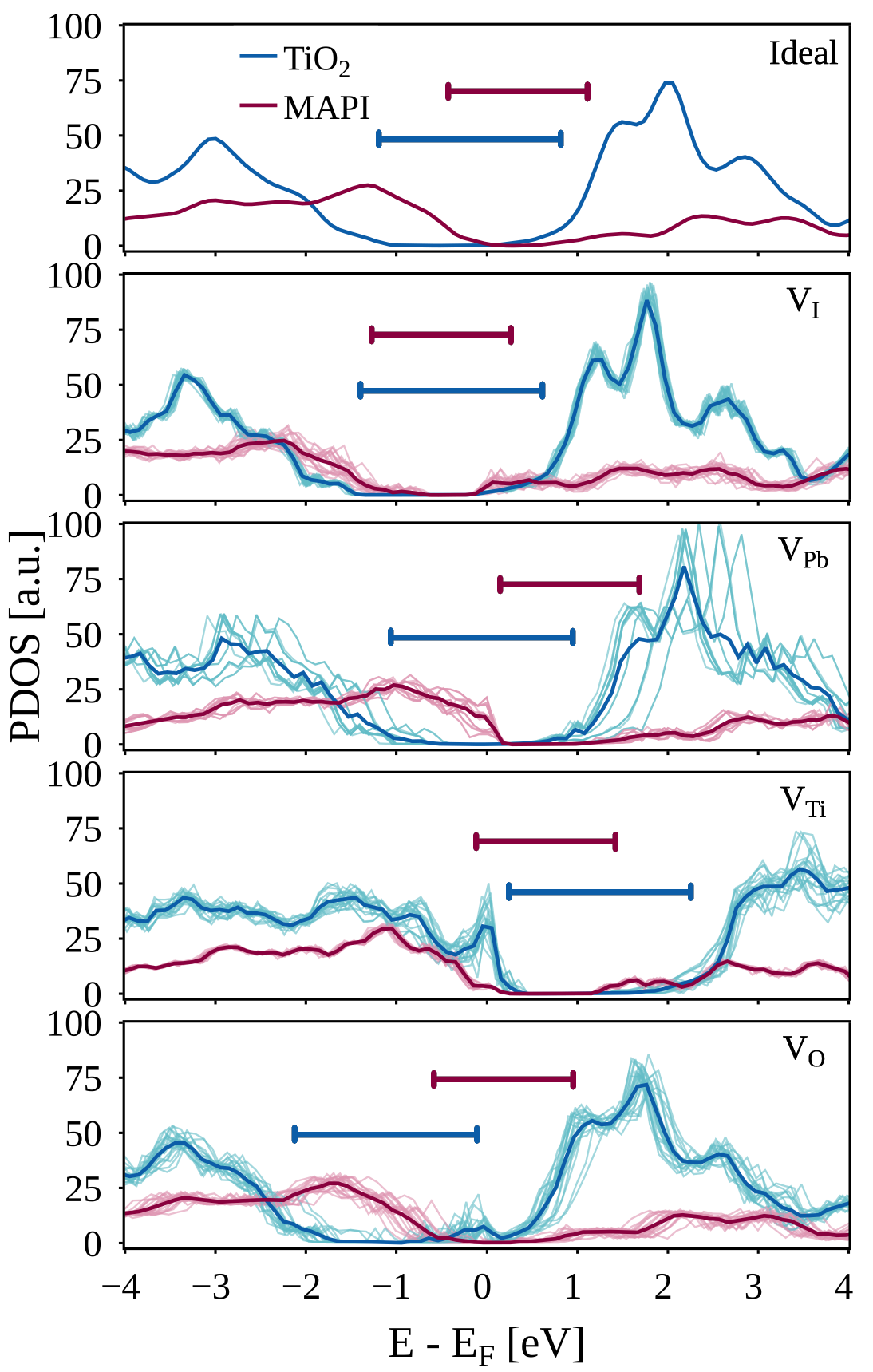}\hspace*{0.5cm}
          \includegraphics[width=0.47\linewidth]{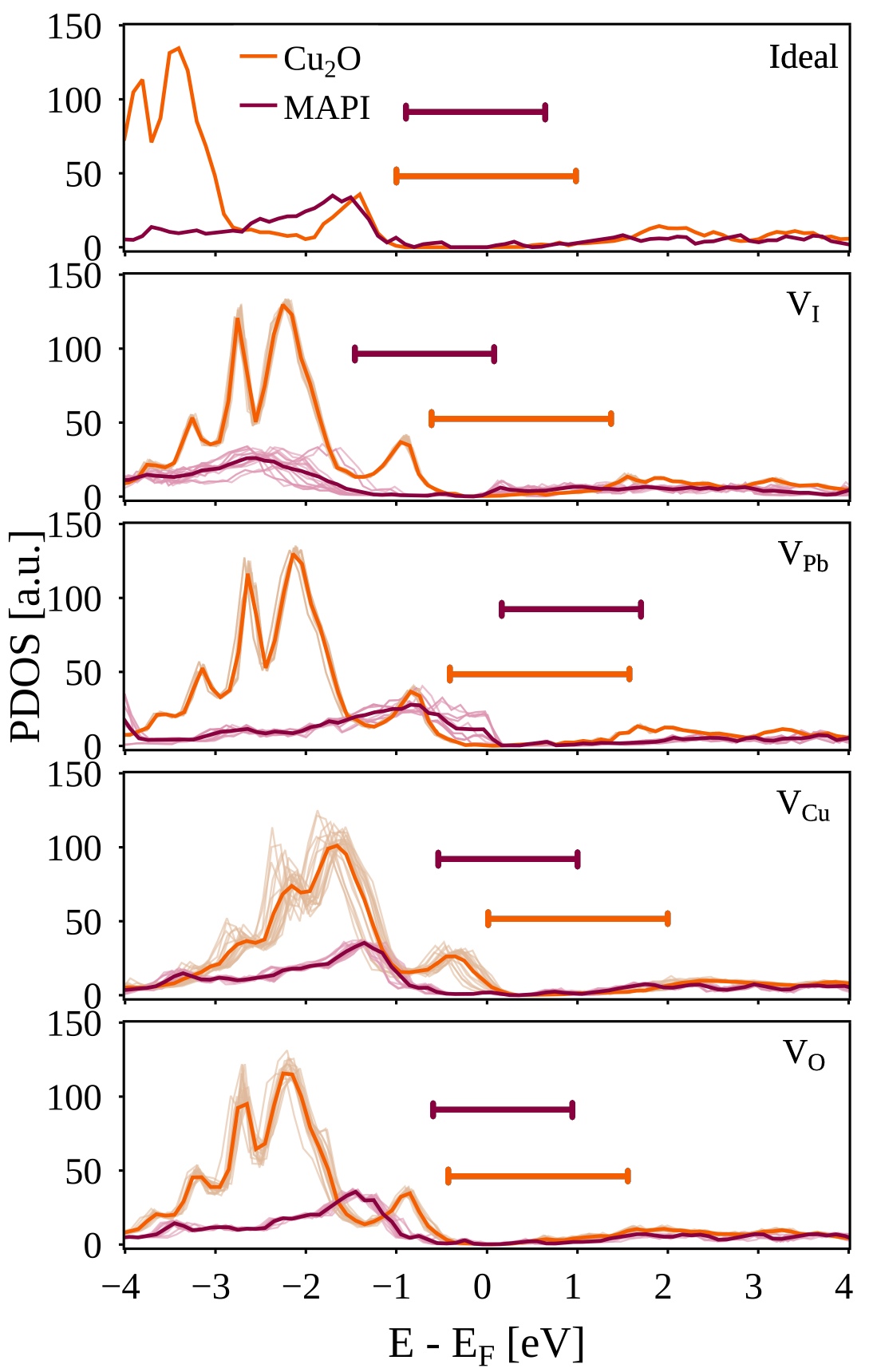}
        \end{center}
	    \caption{Band alignment at MAPI@TiO$_2$ (a) and  MAPI@Cu$_2$O (b) interfaces obtained from the PDOS of the two materials: ideal system (upper plot), followed by structures with $V_{\rm I}$, $V_{\rm Pb}$, $V_{\rm Ti}$/ $V_{\rm Cu}$ and $V_{\rm O}$. A number of 20 disorder realizations are depicted together with the averaged PDOS.}
            \label{MAPI_oxide}
        \end{figure*}

The optimized atomic coordinates of the bulk materials are used to construct the interface models MAPI@TiO$_2$ and MAPI@Cu$_2$O. For  the former interface, a $\pi/4$ rotated orientation of MAPI around $c$-axis is preferred to acquire a better fit. The remaining mismatch was then eliminated by lattice scaling of the absorber. In order to consistently simulate thicker films as in the experimental devices, the first oxide layers were fixed during the relaxation in both interface structures.    

Initially, ideal systems are considered. Following relaxations, the PDOS is determined, summing up contributions from the two materials. Using the procedures outlined in Refs. \cite{doi:10.1021/jz501127k,doi:10.1021/jz501869f,C5CP05466D}, the upper plots of Fig.\ \ref{MAPI_oxide} show the relative shifts of the band gaps, corresponding to the ETL and HTL functionalities. Small band offsets are evidenced for the conduction band for interface $I_1$ and for the valence band in case of interface $I_2$, in accordance with band diagram in Fig.\ \ref{refPSCstruct}.

Introducing vacancies a strong doping effect is observed in the IDLs, with either $n$- or $p$-type character depending on the impurity type. For the oxides, $V_{\rm O}$ create donor centers, while $V_{\rm Ti}$ and $V_{\rm Cu}$ provide acceptors. Similarly, for MAPI, $V_{\rm I}$ induce an $n$-type character, while $V_{\rm Pb}$ introduce a $p$-type character. 
The different disorder realizations can render slightly different outcomes in the PDOS analysis, depending on their relative position and the proximity to the interface. However, the 20 cases considered for each vacancy type present a consistent trend. A careful inspection of Fig.\ \ref{MAPI_oxide}(a) shows that MAPI PDOS is shifted in opposite directions for $V_{\rm I}$ and $V_{\rm Pb}$, followed by a similar effect observed for TiO$_2$ when $V_{\rm Ti}$ or $V_{\rm O}$ are present. High defect concentrations can enhance the band offsets or can even locally reverse the band alignment. The same behavior is evidenced for the MAPI@Cu$_2$O interface in Fig.\ \ref{MAPI_oxide}(b). In addition, the surface terminations of the two slabs can play a role in the charge collection as revealed in a recent study \cite{BAYANI2026109366}.

To further elucidate the role of point defects at the ETL/PRV and PRV/HTL interfaces, a systematic analysis of vacancy formation energies was performed. In particular, iodine and lead vacancies were considered at the MAPI@TiO$_2$ and MAPI@Cu$_2$O interfaces in order to assess their thermodynamic stability and their potential impact on interfacial electronic properties.

The formation energy ($E_f$) of a vacancy defect is defined as:
\begin{equation}
E_f = E_{\mathrm{def}} - E_{\mathrm{ideal}} + \sum_i n_i \mu_i ,
\end{equation}
where $E_{\mathrm{def}}$ and $E_{\mathrm{ideal}}$ denote the total energies of the defective and pristine junction systems, respectively. The term $n_i$ represents the number of atoms of species $i$ removed to create the vacancy ($n_i > 0$), and $\mu_i$ is the corresponding chemical potential. In this work, $i$ refers to iodine and lead atoms.

In order to obtain comparable defect concentrations at the two interfaces, different numbers of iodine vacancies were introduced, reflecting the distinct numbers of iodine atoms in the respective supercells. For the TiO$_2$@MAPI interface, two iodine vacancies were considered, corresponding to a vacancy concentration of approximately 6.3\%. For the Cu$_2$O@MAPI interface, a single iodine vacancy was introduced, resulting in a defect concentration of about 4.5\%. Although the target concentration was approximately 5\%, slight deviations were unavoidable due to the finite size of the systems. Nevertheless, the resulting concentrations remain sufficiently close to enable a meaningful comparison.

At the TiO$_2$@MAPI interface, the average value for the formation energy is $3.87 \pm 0.68$~eV (double vacancies), compared to $3.08 \pm 1.18$~eV (single vacancies) at the Cu$_2$O@MAPI interface. Conversely, single lead vacancies are less likely to form at the Cu$_2$O@MAPI interface, with an average formation energy of $4.40 \pm 1.14$~eV, compared to $2.65 \pm 0.55$~eV at the TiO$_2$@MAPI interface.

These results indicate that iodine and lead vacancies preferentially form at the ETL/PRV interface. Such defects are therefore expected to play a significant role in modifying interfacial band alignment, charge transport, and recombination processes, thereby influencing device stability and performance.

\subsection{Impact of the different defect types on the PSC performance}

The reference solar cell presents a nearly flat band condition for electrons at interface $I_1$ and for holes at interface $I_2$, which indicates a rather unimpeded photogenerated carrier collection. From the simulated J-V characteristics we obtain the following performance parameters: PCE = 19.25\%, FF = 82\%, $V_{\rm oc} = 1.04$ V, $I_{\rm sc} = 22.5$ mA.

    \begin{figure}[t]%
        \begin{center}
          \includegraphics[width=0.85\linewidth]{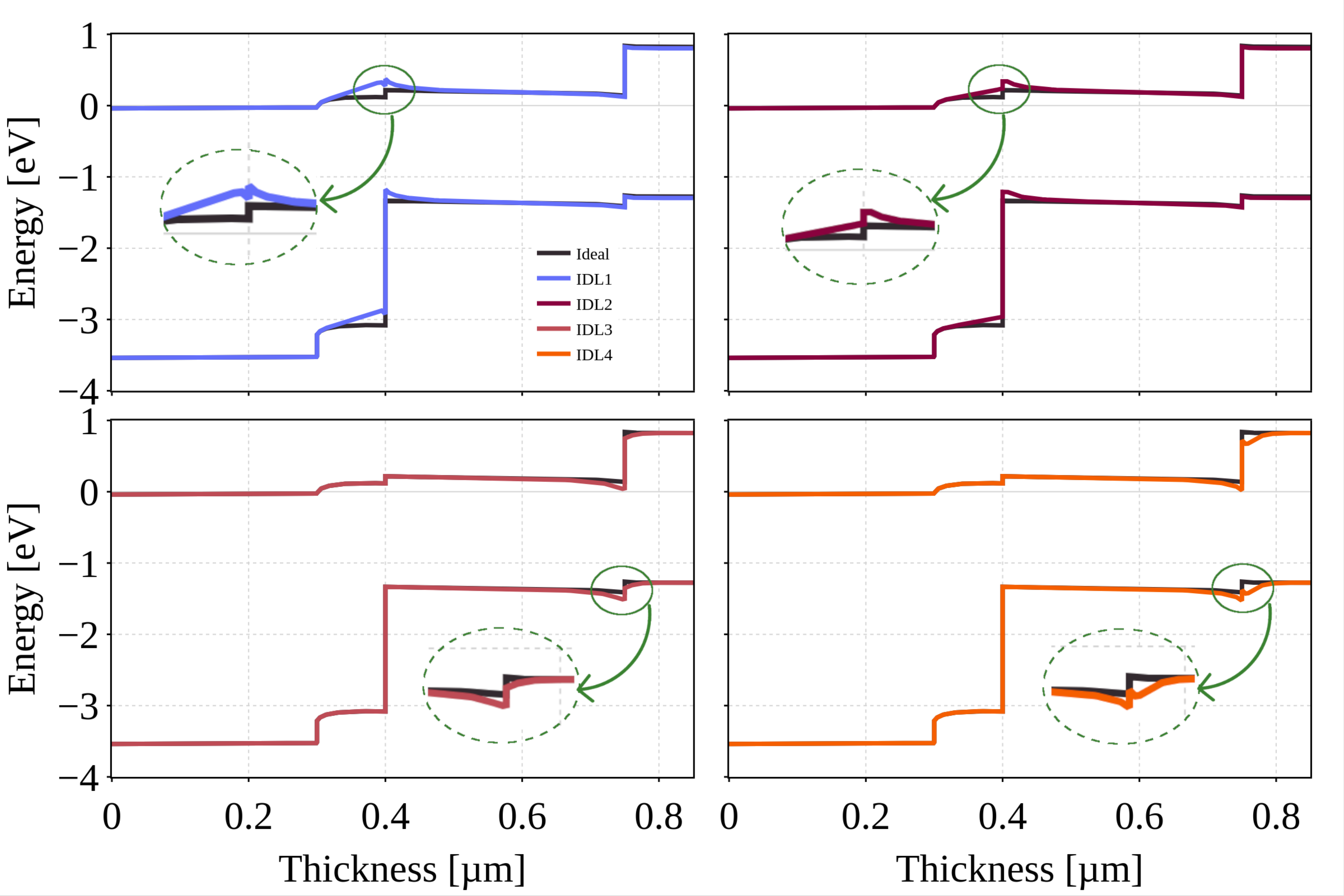}
        \end{center}
	    \caption{Band diagrams obtained by SCAPS simulations: IDL1 and IDL2 have p-type vacancies ($V_{\rm Ti}$ and $V_{\rm Pb}$), while IDL3 and IDL4 have n-type vacancies ($V_{\rm I}$ and $V_{\rm O}$). The small bumps (IDL1,IDL2) and dips (IDL3,IDL4) affect the electron and hole transfer to the ETL and HTL, respectively. The vacancy defect concentration considered here is $10^{18}$ cm$^{-3}$.}
            \label{band_diagram}
        \end{figure}

Vacancy-induced doping can alter significantly the performance of the PSCs and different combinations of defects are possible. However, for the investigated structure, we show that only certain vacancy types have a significant impact. Following the DFT analysis on the band alignment, we expect that the IDLs will create local variations on the band edges near the interfaces. We consider vacancy defects in concentrations up to $5\times 10^{18}$ cm$^{-3}$ of either donor or acceptor type. The band diagrams are presented in Fig.\ \ref{band_diagram}. In the case of $I_1$ interface, $V_{\rm O}$ and $V_{\rm I}$ induce small dips, while $V_{\rm Ti}$ and $V_{\rm Pb}$ produce small bumps. Taking into account the small band offset in the conduction band, $V_{\rm Ti}$ and $V_{\rm Pb}$ found in IDL1 and IDL2, respectively, will have a dominant effect on the electron collection. The holes are blocked at this interface and the large band offset is less affected by the same amount of defects. On the other hand, the small dips induced by $V_{\rm O}$ and $V_{\rm I}$ have a much lesser impact on the electron collection. This image is mirrored at the $I_2$ interface. In this case, the holes are the collected carriers, and they are mostly affected by an $n$-type character of the IDLs. This corresponds to $V_{\rm I}$ and $V_{\rm O}$ in IDL3 and IDL4, respectively. The trapping and collection of photogenerated carriers at the interfaces has been previously discussed in the context of SCAPS simulations in different perovskite solar cell structures and similar spike- or cliff-like features were observed in the band alignment near the interfaces for certain combinations of ETL/perovskite and perovskite/HTL materials \cite{D5SU00526D}. However, our analysis shows the local impact of p- and n-type doping on the band alignment resulting from the interfacial defects.

    \begin{figure*}%
        \begin{center}
          \includegraphics[width=0.85\linewidth]{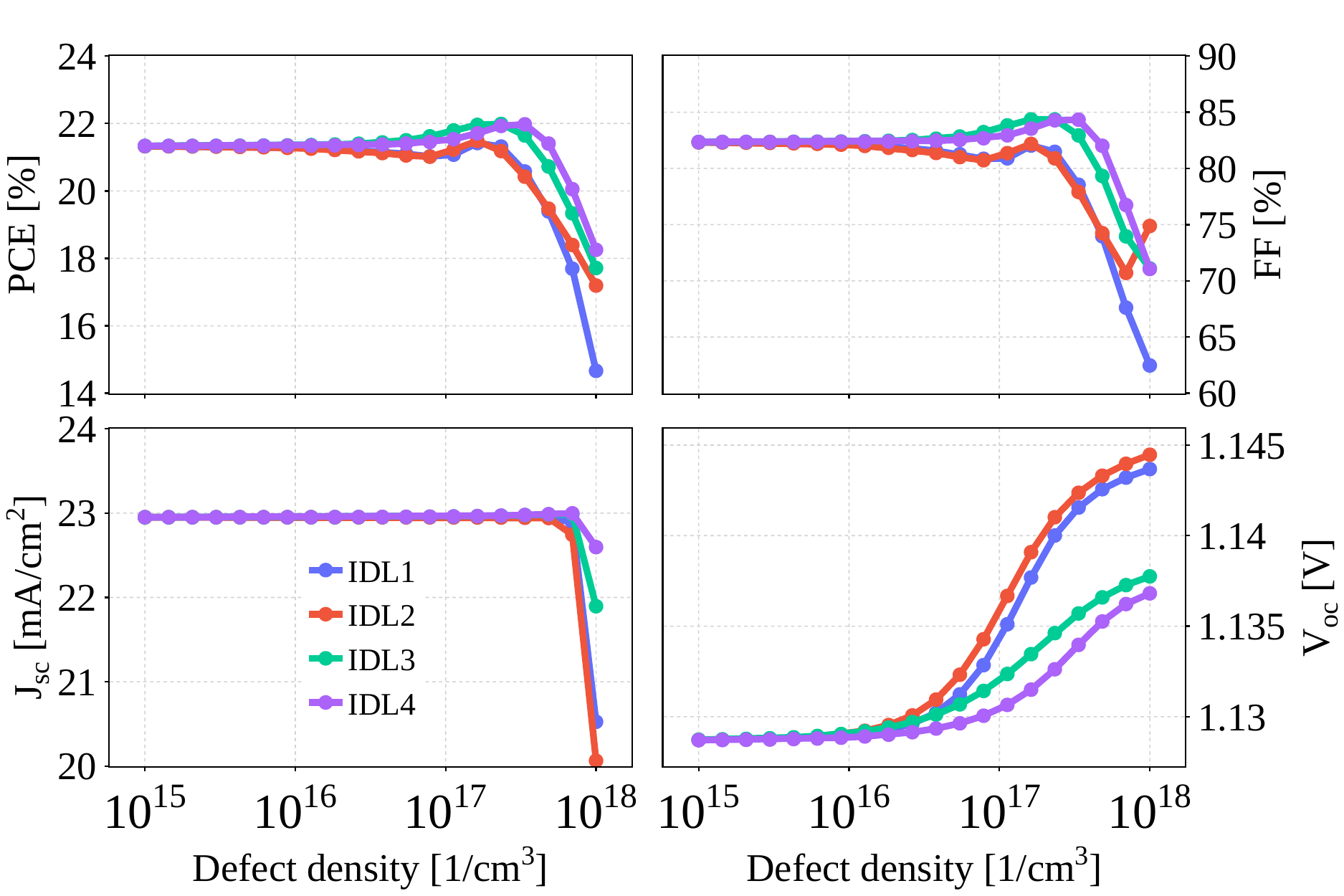}
        \end{center}
	    \caption{Performance of the PSCs with p-type IDL1/IDL2 and n-type IDL3/IDL4, as evidenced by PCE, FF, $I_{\rm sc}$ and $V_{\rm oc}$ as a function of interface defect density.}
            \label{perform}
        \end{figure*}

The performance of the PSCs depending on the defect concentration in the $10^{15}$ - $10^{18}$ cm$^{-3}$ range is shown in Fig.\ \ref{perform}, by monitoring several indicators: PCE, FF, $I_{\rm sc}$ and $V_{\rm oc}$. As the defect-induced features in the conduction and valence bands grow with increasing defect concentration, the PCE drops with 7\%. Within the considered concentration range, interface $I_1$ proves to be more sensitive to defects compared to $I_2$. The same trend is followed by $I_{\rm sc}$ and FF. The open circuit voltage shows a slight non-monotonic increase, which stabilizes at around 1.13-1.14 V. More complex defects that are likely to occur near the interface will further reduce the PCE and further promote long term degradation. Naturally, the sensitivity to defect concentration can be diminished by increasing the CB and VB offsets for $I_1$ and $I_2$, respectively. In this case, the oxide defects would have a lesser impact on the J-V characteristics.

\section{Conclusions}

The performance of PSCs with Cu$_2$O as inorganic HTL in the presence of point defects was investigated using {\it ab initio} calculations and SCAPS simulations. In the first step, the band alignment in TiO$_2$/MAPI/Cu$_2$O based PSCs was determined by DFT calculations. Different types of vacancies induce either $n$- or $p$-type doping in the ETL, HTL and MAPI layers, which impact the overall solar cell performance. It was established that in a mostly flat band condition for electrons at the ETL-PRV interface and for holes at PRV-HTL interface, the dominant vacancies are $V_{\rm Ti}$ and $V_{\rm O}$ in the two oxide layers, TiO$_2$ and Cu$_2$O, respectively, while for MAPI $V_{\rm I}$ impacts the interface with the ETL and $V_{\rm Pb}$ influence the interface with the HTL. Therefore, titanium vacancies in TiO$_2$ and oxygen vacancies in Cu$_2$O may play a significant role in nearly flat-band PSCs and could represent important targets for passivation strategies. The SCAPS simulations confirm this perspective, by analyzing the PCE dependence on the defect concentration, type and layer localization. This framework would enable in future studies additional quantitative input from DFT into SCAPS simulations, in form of trap states distribution and donor/acceptor like character of more complex defects, beyond the point-defects considered here. To conclude, we performed a comprehensive study on the impact of vacancy defects and provided clues to mitigate potential performance decrease by unintentional defect-induced doping of the active layers. \\

%
%

\section*{Data availability statement}

All data generated or analyzed during this study are included in this published manuscript.

\section*{Acknowledgment}

The research leading to these results has received funding from the EEA Grants 2014–2021, under Project contract no. 36/2021 (project code: EEA-RO-NO-2018-0106). Additional funds from Core Programs PN23210204 and PN23080303 are also acknowledged. This work benefited from services and resources provided by EGI, with the dedicated support of CLOUDIFIN.





\bibliographystyle{iopart-num}
\bibliography{manuscript}

\providecommand{\newblock}{}
\begin{thebibliography}{10}
\expandafter\ifx\csname url\endcsname\relax
  \def\url#1{{\tt #1}}\fi
\expandafter\ifx\csname urlprefix\endcsname\relax\def\urlprefix{URL }\fi
\providecommand{\eprint}[2][]{\url{#2}}

\bibitem{nrel}
 {\em Best Research Cells Efficiencies from NREL (2026),
  https://www.nrel.gov/pv/cell-efficiency.html\/}

\bibitem{KHATOON2023437}
Khatoon S, {Kumar Yadav} S, Chakravorty V, Singh J, {Bahadur Singh} R, Hasnain
  M~S and Hasnain S~M 2023 {\em Materials Science for Energy Technologies\/}
  {\bf 6} 437--459

\bibitem{Tao2023}
Tao S, Polavarapu L and Vivo P 2023 {\em Scientific Reports\/} {\bf 13} 4370

\bibitem{Lee_2022}
Lee H~J, Seo Y~H, Kim S~S and Na S~I 2022 {\em Semiconductor Science and
  Technology\/} {\bf 37} 045007

\bibitem{SIN2023112054}
Sin J, Kim H, Kim M, Kim M, Shin J, Hong J and Yang J 2023 {\em Solar Energy
  Materials and Solar Cells\/} {\bf 250} 112054

\bibitem{JIAO2023101158}
Jiao J, Yang C, Wang Z, Yan C and Fang C 2023 {\em Results in Engineering\/}
  {\bf 18} 101158

\bibitem{en14175431}
Filipoiu N, Mitran T~L, Anghel D~V, Florea M, Pintilie I, Manolescu A and
  Nemnes G~A 2021 {\em Energies\/} {\bf 14} 5431

\bibitem{doi:10.1021/acs.jpclett.6b02375}
Besleaga C, Abramiuc L~E, Stancu V, Tomulescu A~G, Sima M, Trinca L, Plugaru N,
  Pintilie L, Nemnes G~A, Iliescu M, Svavarsson H~G, Manolescu A and Pintilie I
  2016 {\em The Journal of Physical Chemistry Letters\/} {\bf 7} 5168--5175

\bibitem{https://doi.org/10.1002/nano.202000238}
Wang Q, Lin Z, Su J, Hu Z, Chang J and Hao Y 2021 {\em Nano Select\/} {\bf 2}
  1055--1080

\bibitem{SAJID2023101378}
Sajid S, Alzahmi S, Salem I~B, Park J and Obaidat I~M 2023 {\em Materials Today
  Energy\/} {\bf 37} 101378

\bibitem{https://doi.org/10.1002/nano.202000200}
Arumugam G~M, Karunakaran S~K, Liu C, Zhang C, Guo F, Wu S and Mai Y 2021 {\em
  Nano Select\/} {\bf 2} 1081--1116

\bibitem{electronics14224428}
Rijal B, Alzoubi K~M, Chen Z, Latif~Khammash A, Lu S and Dhakal T~P 2025 {\em
  Electronics\/} {\bf 14} 4428

\bibitem{CAO2022420}
Cao W, Hu Z, Lin Z, Guo X, Su J, Chang J and Hao Y 2022 {\em Journal of Energy
  Chemistry\/} {\bf 68} 420--438

\bibitem{C5CC05205J}
Frolova L~A, Dremova N~N and Troshin P~A 2015 {\em Chem. Commun.\/} {\bf 51}
  14917--14920

\bibitem{Srivastava2023}
Srivastava S, Ranjan S, Yadav L, Sharma T, Choudhary S, Agarwal D, Singh A,
  Satapathi S, Gupta R~K, Garg A and Nalwa K~S 2023 {\em Communications
  Materials\/} {\bf 4} 52

\bibitem{Euvrard2021}
Euvrard J, Yan Y and Mitzi D~B 2021 {\em Nature Reviews Materials\/} {\bf 6}
  531--549

\bibitem{CHOUHAN2018150}
Chouhan A~S, Jasti N~P and Avasthi S 2018 {\em Materials Letters\/} {\bf 221}
  150--153

\bibitem{doi:10.1021/jz501127k}
Mosconi E, Ronca E and De~Angelis F 2014 {\em The Journal of Physical Chemistry
  Letters\/} {\bf 5} 2619--2625

\bibitem{doi:10.1021/jz501869f}
Colella S, Mosconi E, Pellegrino G, Alberti A, Guerra V~L~P, Masi S, Listorti
  A, Rizzo A, Condorelli G~G, De~Angelis F and Gigli G 2014 {\em The Journal of
  Physical Chemistry Letters\/} {\bf 5} 3532--3538

\bibitem{C5CP05466D}
Nemnes G~A, Goehry C, Mitran T~L, Nicolaev A, Ion L, Antohe S, Plugaru N and
  Manolescu A 2015 {\em Phys. Chem. Chem. Phys.\/} {\bf 17} 30417--30423

\bibitem{Geng2016}
Geng W, Tong C~J, Liu J, Zhu W, Lau W~M and Liu L~M 2016 {\em Scientific
  Reports\/} {\bf 6} 20131

\bibitem{YANG2018394}
Yang Z, Wang Y and Liu Y 2018 {\em Applied Surface Science\/} {\bf 441}
  394--400

\bibitem{doi:10.1021/acs.jpclett.3c03536}
Chen X, Pasanen H~P, Khan R, Tkachenko N~V, Janáky C and Samu G~F 2024 {\em
  The Journal of Physical Chemistry Letters\/} {\bf 15} 2057--2065

\bibitem{NICOLAEV2016202}
Nicolaev A, Mitran T, Iftimie S and Nemnes G 2016 {\em Solar Energy Materials
  and Solar Cells\/} {\bf 158} 202--208

\bibitem{Si2017}
Si F, Hu W, Tang F, Cheng Y and Xue H 2017 {\em Journal of Materials Science\/}
  {\bf 52} 13841--13851

\bibitem{Sultana2019}
Sultana N, Al~Amin A, Metin D~Z and Gaston N 2019 {\em Journal of Materials
  Science\/} {\bf 54} 13594--13608

\bibitem{MOHAMMADI2025118454}
Mohammadi M~R and Saremi F 2025 {\em Materials Science and Engineering: B\/}
  {\bf 320} 118454

\bibitem{doi:10.1021/acsami.0c11187}
Castellanos-Águila J~E, Lodeiro L, Menéndez-Proupin E, Montero-Alejo A~L,
  Palacios P, Conesa J~C and Wahnón P 2020 {\em ACS Applied Materials \&
  Interfaces\/} {\bf 12} 44648--44657

\bibitem{MITRAN2023153}
Mitran T~L, Brophy R~E, Cuzminschi M, Filipoiu N, Kateb M, Pintilie I,
  Manolescu A and Nemnes G~A 2023 Chapter 5 - ab initio studies on perovskites
  {\em Low-Dimensional Halide Perovskites\/} ed Zhan Y, Khalid M, Vivo P and
  Arshid N (Elsevier) pp 153--185

\bibitem{BURGELMAN2000527}
Burgelman M, Nollet P and Degrave S 2000 {\em Thin Solid Films\/} {\bf 361-362}
  527--532

\bibitem{MINEMOTO20141428}
Minemoto T and Murata M 2014 {\em Current Applied Physics\/} {\bf 14}
  1428--1433

\bibitem{Goje_2023}
Goje A~A, Ludin N~A, Teridi M~A~M, Syafiq U, Ibrahim M~A, Nawab F and Syakirin
  A~A 2023 {\em IOP Conference Series: Materials Science and Engineering\/}
  {\bf 1278} 012004

\bibitem{D2RA06734J}
Hossain M~K, Arnab A~A, Das R~C, Hossain K~M, Rubel M~H~K, Rahman M~F,
  Bencherif H, Emetere M~E, Mohammed M~K~A and Pandey R 2022 {\em RSC Adv.\/}
  {\bf 12} 35002--35025

\bibitem{doi:10.1021/acs.energyfuels.3c00035}
Hossain M~K, Mohammed M~K~A, Pandey R, Arnab A~A, Rubel M~H~K, Hossain K~M, Ali
  M~H, Rahman M~F, Bencherif H, Madan J, Islam M~R, Samajdar D~P and Bhattarai
  S 2023 {\em Energy \& Fuels\/} {\bf 37} 6078--6098

\bibitem{Baro2023}
Baro M and Borgohain P 2023 {\em Journal of Electronic Materials\/} {\bf 52}
  7623--7644

\bibitem{inorganics11010003}
Pochont N~R and Sekhar Y~R 2023 {\em Inorganics\/} {\bf 11} 3

\bibitem{nano11051218}
Samiul~Islam M, Sobayel K, Al-Kahtani A, Islam M~A, Muhammad G, Amin N,
  Shahiduzzaman M and Akhtaruzzaman M 2021 {\em Nanomaterials\/} {\bf 11} 1218

\bibitem{HUNDE2023105420}
Hunde B~R and Woldeyohannes A~D 2023 {\em Materials Today Communications\/}
  {\bf 34} 105420

\bibitem{MENG2022112427}
Meng X, Tang T, Zhang R, Liu K, Li W, Yang L, Song Y, Ma X, Cheng Z and Wu J
  2022 {\em Optical Materials\/} {\bf 128} 112427

\bibitem{ZHANG2023128084}
Zhang Y, Meng X, Liu X, Zhou F, Yang W, Fan Y, He P, Wu J, Wang H and Cheng Y
  2023 {\em Materials Chemistry and Physics\/} {\bf 306} 128084

\bibitem{D5SU00526D}
Araujo V~H~D, Nogueira A~F, Tristao J~C and Santos L~J~d 2025 {\em RSC
  Sustainability\/} {\bf 3} 4314--4335

\bibitem{0953-8984-14-11-302}
Soler J~M, Artacho E, Gale J~D, Garcia A, Junquera J, Ordejon P and
  Sanchez-Portal D 2002 {\em Journal of Physics: Condensed Matter\/} {\bf 14}
  2745

\bibitem{PhysRevLett.45.566}
Ceperley D~M and Alder B~J 1980 {\em Phys. Rev. Lett.\/} {\bf 45} 566--569

\bibitem{PhysRevB.43.1993}
Troullier N and Martins J~L 1991 {\em Phys. Rev. B\/} {\bf 43} 1993--2006

\bibitem{YAN201434}
Yan X~G, Xu L, Huang W~Q, Huang G~F, Yang Z~M, Zhan S~Q and Long J~P 2014 {\em
  Materials Science in Semiconductor Processing\/} {\bf 23} 34--41

\bibitem{abia2022crystal}
Abia C, L{\'o}pez C~A, Ca{\~n}adillas-Delgado L, Fern{\'a}ndez-Diaz M~T and
  Alonso J~A 2022 {\em Scientific Reports\/} {\bf 12} 18647

\bibitem{whitfield2016structures}
Whitfield P, Herron N, Guise W, Page K, Cheng Y, Milas I and Crawford M 2016
  {\em Scientific reports\/} {\bf 6} 35685

\bibitem{zerothi_sisl}
Papior N 2024 sisl: v0.14.3
  \urlprefix\url{https://doi.org/10.5281/zenodo.597181}

\bibitem{bransky1969hall}
Bransky I and Tannhauser D 1969 {\em Solid State Communications\/} {\bf 7}
  245--248

\bibitem{ait2022comparison}
Ait-Wahmane Y, Mouhib H, Ydir B, Ait~Hssi A, Atourki L, Ihlal A and Bouabid K
  2022 {\em Materials Today: Proceedings\/} {\bf 52} 166--171

\bibitem{bansal2016evaluation}
Bansal S and Aryal P 2016 Evaluation of new materials for electron and hole
  transport layers in perovskite-based solar cells through scaps-1d simulations
  {\em 2016 IEEE 43rd photovoltaic specialists conference (PVSC)\/} (IEEE) pp
  0747--0750

\bibitem{tan2016controllable}
Tan K, Lin P, Wang G, Liu Y, Xu Z and Lin Y 2016 {\em Solid-State
  Electronics\/} {\bf 126} 75--80

\bibitem{abena2022numerical}
Abena A~N, Ngoupo A~T, Abega F~A and Ndjaka J 2022 {\em Chinese Journal of
  Physics\/} {\bf 76} 94--109

\bibitem{sajid2024boosting}
Sajid M, Ali K, Shah H~S, Shahid A, Bakar S~A, Yousaf M and Lee S~L 2024 {\em
  Journal of Optics\/}  1--12

\bibitem{sekkat2021open}
Sekkat A, Nguyen V~H, Masse~de La~Huerta C~A, Rapenne L, Bellet D,
  Kaminski-Cachopo A, Chichignoud G and Mu{\~n}oz-Rojas D 2021 {\em
  Communications materials\/} {\bf 2} 78

\bibitem{pan2024high}
Pan L, Dai L, Burton O~J, Chen L, Andrei V, Zhang Y, Ren D, Cheng J, Wu L,
  Frohna K {\em et~al.\/} 2024 {\em Nature\/} {\bf 628} 765--770

\bibitem{MA2021167107}
Ma Z, Ren F, Yang Z and Volinsky A~A 2021 {\em Optik\/} {\bf 241} 167107

\bibitem{PhysRevB.87.115111}
Heinemann M, Eifert B and Heiliger C 2013 {\em Phys. Rev. B\/} {\bf 87} 115111

\bibitem{10.1063/1.4919323}
Murali D~S, Kumar S, Choudhary R~J, Wadikar A~D, Jain M~K and Subrahmanyam A
  2015 {\em AIP Advances\/} {\bf 5} 047143

\bibitem{C4TA05033A}
Yin W~J, Yang J~H, Kang J, Yan Y and Wei S~H 2015 {\em J. Mater. Chem. A\/}
  {\bf 3} 8926--8942

\bibitem{BAYANI2026109366}
Bayani A 2026 {\em Results in Engineering\/} {\bf 29} 109366

\end{thebibliography}

\end{document}